# The Building Blocks of Economic Complexity

*César A. Hidalgo[1†], Ricardo Hausmann[1]*

[1] *Center for International Development and Harvard Kennedy School at Harvard University*
† *To whom correspondence should be addressed* cesar_hidalgo@ksg.harvard.edu

## Abstract:

For Adam Smith, wealth was related to the division of labor. As people and firms specialize in different activities, economic efficiency increases, suggesting that development is associated with an increase in the number of individual activities and with the complexity that emerges from the interactions between them. Here we develop a view of economic growth and development that gives a central role to the complexity of a country's economy by interpreting trade data as a bipartite network in which countries are connected to the products they export, and show that it is possible to quantify the complexity of a country's economy by characterizing the structure of this network. Furthermore, we show that the measures of complexity we derive are correlated with a country's level of income, and that deviations from this relationship are predictive of future growth. This suggests that countries tend to converge to the level of income dictated by the complexity of their productive structures, indicating that development efforts should focus on generating the conditions that would allow complexity to emerge in order to generate sustained growth and prosperity.

# Introduction

For Adam Smith, the secret to the wealth of nations was related to the division of labor. As people and firms specialize in different activities, economic efficiency increases. This division of labor, however, is limited by the extent of the market: the bigger the market, the more its participants can specialize and the deeper the division of labor that can be achieved. This suggests that wealth and development are related to the complexity that emerges from the interactions between the increasing number of individual activities that conform an economy [1,2,3].

Now, if all countries are connected to each other through a global market for inputs and outputs so that they can exploit a division of labor at the global scale, why have differences in Gross Domestic Product (GDP) per capita exploded over the past two centuries?[*] [4,5] One possible answer is that some of the individual activities that arise from the division of labor described above cannot be imported, such as property rights, regulation, infrastructure, specific labor skills, etc., and so countries need to have them locally available in order to produce. Hence, the productivity of a country resides in the diversity of its available non-tradable capabilities, and therefore, cross-country differences in income can be explained by differences in economic complexity, as measured by the diversity of capabilities present in a country and their interactions.

During the last twenty years, models of economic growth have often included the assumption that the variety of inputs that go into the production of the goods produced by a country affects that country's overall productivity [3,6]. However, there have been very few attempts to bring this intuition to the data. In fact, the most frequently cited surveys of the empirical literature do not incorporate a single reference to any measure of diversity of inputs or complexity [7].

---

[*] In [4] Maddison presents GDP per capita measures for 60 countries since 1820. In that year, the ratio of the 95th to the 5th percentile was 3.18 but it increased to 17.82 by the year 2000. Today, the US GDP per capita is over 60 times higher than Malawi's.

We can create indirect measures of the capabilities available in a country by thinking of each one of these capabilities as a building block or Lego piece. In this analogy, a product is equivalent to a Lego model, and a country is equivalent to a bucket of Legos. Countries will be able to make products for which they have all the necessary capabilities, just like a child is able to produce a Lego model if the child's bucket contains all the necessary Lego pieces. Using this analogy, the question of economic complexity is equivalent to asking whether we can infer properties such as the diversity and exclusivity of the Lego pieces inside a child's bucket by looking only at the models that a group of children, each with a different bucket of Legos, can make. Here we show that this is possible if we interpret data connecting countries to the products they export as a bipartite network and assume that this network is the result of a larger, tripartite network, connecting countries to the capabilities they have and products to the capabilities they require (Fig 1a). Hence, connections between countries and products signal the availability of capabilities in a country just like the creation of a model by a child signals the availability of a specific set of Lego pieces.

Note that this interpretation says nothing of the processes whereby countries accumulate capabilities and the characteristics of an economy that might affect them. It just attempts to develop measures of the complexity of a country's economy at a point in time. However, the approach presented here can be seen as a building block of a theory that accounts for the process by which countries accumulate capabilities. A detailed analysis of capability accumulation is beyond the scope of this paper but the implications of our approach will be discussed in the concluding section, after we have presented our results.

In this paper we develop a method to characterize the structure of bipartite networks, which we call the Method of Reflections, and apply it to trade data to illustrate how it can be used to extract relevant information about the availability of capabilities in a country. We interpret the variables produced by the Method of Reflections as indicators of economic complexity and show that the complexity of a country's economy is correlated with income and that deviations from this relationship are predictive of future growth, suggesting that countries tend to approach the level of income associated with the capability set available in them. We validate our measures of the capabilities available in a country by introducing a model and by showing empirically that our metrics are strongly correlated with the *diversity of the labor inputs* used in the production of a country's goods, assuming that they use the same labor inputs per product as in the United States. Finally, we show that the level of complexity of a country's economy predicts the types of products that countries will be able to develop in the future, suggesting that the new products that a country develops depend substantially on the capabilities already available in that country.

## Methods

We look at country product associations by using international trade data with products disaggregated according to three alternative data sources and classifications: first, the Standard International Trade Classification SITC revision 4 at the four-digit level [†] [8], second, the COMTRADE Harmonized System at the 4-digit level and third, the North American Industry Classification System (NAICS) at the 6 digit level (SM Section 1). We interpret these data as bipartite networks in which countries are connected to the products they export (Fig 1b). Mathematically, we represent this network using the adjacency matrix $M_{cp}$, where $M_{cp}=1$ if country $c$ is a significant exporter of product $p$ and 0 otherwise. We consider country $c$ to be a significant exporter of product $p$ if its Revealed Comparative Advantage (the share of product $p$ in the export basket of country $c$ to the share of product $p$ in world trade) is greater than some threshold value, which we take as 1 in this exercise ($RCA_{cp} \geq 1$) (see SM Section 2).

---

[†] The data is available at http://www.nber.org/data  http://cid.econ.udavis.edu/data/undata/undata.html and http://www.chidalgo.com/productspace/data.html

**Method of Reflections**

We characterize countries and products by introducing a family of variables capturing the structure of the network defined by $M_{cp}$ (SM Section 3). Because of the symmetry of the bipartite network, we refer to this technique as the "Method of Reflections", as the method produces a symmetric set of variables for the two types of nodes in the network (countries and products).

The Method of Reflections consists of iteratively calculating the average value of the previous-level properties of a node's neighbors and is defined as the set of observables:

$$k_{c,N} = \frac{1}{k_{c,0}} \sum_p M_{cp} k_{p,N-1}, \qquad (1)$$

$$k_{p,N} = \frac{1}{k_{p,0}} \sum_c M_{cp} k_{c,N-1}. \qquad (2)$$

for $N \geq 1$. With initial conditions given by the degree, or number of links, of countries and products:

$$k_{c,0} = \sum_p M_{cp}, \qquad (3)$$

$$k_{p,0} = \sum_c M_{cp}. \qquad (4)$$

$k_{c,0}$ and $k_{p,0}$ represent, respectively, the observed levels of diversification of a country (the number of products exported by that country), and the ubiquity of a product (the number of countries exporting that product). Hence, we characterize each country through the vector $\vec{k}_c = (k_{c,0}, k_{c,1}, k_{c,2}, \ldots, k_{c,N})$ and each product by the vector $\vec{k}_p = (k_{p,0}, k_{p,1}, k_{p,2}, \ldots, k_{p,N})$.

For countries, even variables ($k_{c,0}, k_{c,2}, k_{c,4}, \ldots$) are generalized measures of diversification, whereas odd variables ($k_{c,1}, k_{c,3}, k_{c,5}, \ldots$) are generalized measures of the ubiquity of their exports. For products, even variables are related to their ubiquity and the ubiquity of other related products, whereas odd variables are related to the diversification of countries exporting those products. In network terms, $k_{c,1}$ and $k_{p,1}$ are known as the average nearest neighbor degree [9,10]. Higher order variables, however, ($N>1$) can be interpreted as a linear combination of the properties of all the nodes in the network with coefficients given by the probability that a random walker that started at a given node ends up at another node after $N$ steps (see SM Section 4).

## Results

We can begin understanding the type of information about countries captured by the Method of Reflections by looking at where countries are located in the space defined by the first two sets of variables produced by our method: $k_{c,0}$ and $k_{c,1}$. Fig 1c shows that there is a strong negative correlation between $k_{c,0}$ and $k_{c,1}$ [9,11], meaning that diversified countries tend to export less ubiquitous products. Deviations from this behavior, however, are informative. For example, while Malaysia and Pakistan export the same number of products, the products exported by Malaysia ($k_{MYS,0}=104$, $k_{MYS,1}=18$) are exported by fewer countries than those exported by Pakistan ($k_{PAK,0}$=104, $k_{PAK,1}$=27.5). Combining this fact with our third level of analysis, we see that Malaysian products are exported by more diversified countries than the exports of Pakistan ($k_{MYS,2}$=163 $k_{PAK,2}$=142, SM Section 8). This suggests that the productive structure of Malaysia is more complex than that of Pakistan, due, as we will show shortly, to a larger number of capabilities available in Malaysia than in Pakistan.

In the Supplementary Material we show that the negative relationship presented in the $k_{c,0}$-$k_{c,1}$ diagram is not a consequence of variations in the level of diversification of countries and in the ubiquity of products. We prove this by creating four null models [10] that control, with increasing stringency, for the diversification of countries and the ubiquity of products and show that these distributions, per se, are not responsible for the negative relationship observed in the data (see SM section 6).

**A minimalistic model**

We show that the location of countries in the $k_{c,0}$-$k_{c,1}$ diagram is informative about the capabilities available in a country by introducing a simple model based on the assumption that *country c will be able to produce product p if it has all the required capabilities* (Fig 2a).

We implement this model by considering a fixed number of capabilities in each country and represent this by using a matrix $C_{ca}$, that is equal to 1 if country $c$ has capability $a$ and 0 otherwise. We represent the relationship between capabilities and the products that require them by a matrix $\Pi_{pa}$ whose elements are equal to 1 if product $p$ requires capability $a$ and 0 otherwise.

Using the notation introduced above, together with our only assumption, we can model the structure of the $M_{cp}$ matrix as:

$$M_{cp} = 1 \text{ if } \sum_a \Pi_{pa} = \sum_a \Pi_{pa} C_{ca} \text{ and } M_{cp} = 0 \text{ otherwise} \qquad (5)$$

The simplest implementation of this model is to consider $C_{ca}=1$ with probability $r$ and $0$ with probability $1-r$ and $\Pi_{pa}=1$ with probability $q$ and $0$ with probability $1-q$. An emergent property of the matrix resulting from this model is that the average ubiquity of a country's products tends to decrease with its level of diversification for a wide range of parameters (Fig 2b). We interpret this negative correlation by considering that countries with many capabilities will be more diversified, since they can produce a wider set of products, and that because they can make products requiring many capabilities, few other countries will have all the requisite capabilities to make them, hence diversified countries will be able to make less ubiquitous products.

The model allows us to test directly whether given this set of assumptions we should expect countries with more capabilities to be more diversified and produce less ubiquitous products. Figure 2c shows that, in the model, the diversity of a country increases with the number of capabilities it posses, whereas the ubiquity of a country's products is a decreasing function of the number of capabilities available in that country, providing further theoretical evidence that $\bar{k}_c$ captures information on the availability of capabilities in a country, and therefore, about the complexity of its economy.

**A direct measurement of a subset of capabilities**

We provide empirical evidence that the method of reflections extracts information that is related to the capabilities available in a country by looking at a measurable subset of the capabilities required by products. Figure 2d shows the average number of different employment categories required by products exported by countries versus $k_{c,0}$, $k_{c,1}$ and $k_{c,2}$. We measure the number of employment categories that go into a product by using the data of the US Bureau of Labor Statistics (see SM Section 1). This data should play against us, as we are disregarding the fact that other countries may use different technologies to produce goods that are similarly classified[‡]. Despite this, we find a strong positive correlation between the average number of employment categories going into the export basket of countries and our family of measures of diversification ($k_{c,0}, k_{c,2}, k_{c,4}, \ldots, k_{c,2N}$). We also find a negative correlation between the average number of employment categories and measures of the ubiquity of products made by a country ($k_{c,1}, k_{c,3}, k_{c,5}, \ldots, k_{c,2N+1}$) (Fig 2d). This shows that more diversified countries indeed produce more complex products, in the sense that they require a wider combination of human capabilities, and that $\vec{k}_c$ is able to capture this information.

**The Complexity of the Productive Structure, Income and Growth**

We show that the information extracted by the method of reflections is connected to income by looking at the first three measures of diversification of a country ($k_{c,0}, k_{c,2}, k_{c,4}$) versus GDP per-capita adjusted for Purchasing Power Parity (PPP) (Fig 3 a-c). To make these three different measures comparable we have normalized them by subtracting their respective means ($<k_N>$) and dividing them by their respective standard deviations (stdev($k_N$)). As we iterate the method the relative ranking of countries defined by these variables shifts (Fig 3d Fig S14), making our measures of diversification and ubiquity increasingly more correlated with income (Fig 3e, SM Section 11). This can be illustrated by looking at the position, in the $k_N$–GDP diagrams, of three countries that exported a similar number of products in the year 2000, albeit having large differences in income (Pakistan (PAK), Chile (CHL) and Singapore (SGP) Fig 3 a-c). Higher reflections of our method are able to correctly differentiate the income level of these countries because they incorporate information about the ubiquity of the products they export and about the diversification of other countries connected indirectly to them in $M_{cp}$, altering their relative rankings (Fig 3d Fig S14). For example, $k_{c,2}$ is able to correctly separate Singapore, Chile and Pakistan, because it considers that in the bipartite network Singapore is connected to diversified countries mainly through non-ubiquitous products, signaling the availability in Singapore of capabilities that are required to produce goods in diversified countries. In contrast, Pakistan is connected mostly to poorly diversified countries, and most of its connections are through ubiquitous products, indicating that Pakistan has capabilities that are available in most countries and that its relatively high level of diversification is probably due to its relatively large population, rather than to the complexity of its productive structure. Indeed, we find the method of reflections to be an accurate way to control for a country's population, as correlations between $\vec{k}_c$ and population decrease rapidly as we iterate the method (see SM Section 11), while correlations between $\vec{k}_c$ and GDP increase as we iterate the method. This is another piece of evidence suggesting that the information captured by our method is related to factors that affect the ability to generate per capita income.

---

[‡] Indeed, it is common for poorer countries to exchange labor for capital. For example, building a road in the US is done by a relatively small team of workers, each of them specialized to operate a different machine or technique, while more modest economies will tend to use more workers, yet less specialized ones, as the relative cost of machines to labor is larger in poorer economies. Hence we should expect poor countries to use less labor inputs in the production of products than what would be reported from US labor data, accentuating the effect presented in Fig 2d.

Deviations from the correlation between $\vec{k}_c$ and income are good predictors of future growth, indicating that countries tend to approach the levels of income that correspond to their measured complexity. We show this by regressing the rate of growth of income per capita on successive generations of our measures of economic complexity (i.e. $k_{c,0}, k_{c,1}$ or $k_{c,10}, k_{c,11}$) and on a country's initial level of income

$$\log\left(\frac{GDP(t+\Delta t)}{GDP(t)}\right) = a + b_1 GDP(t) + b_2 k_{c,N}(t) + b_3 k_{c,N+1}(t), \quad (6)$$

finding that successive generations of the variables constructed in the previous section are increasingly good predictors of growth. On section 13 of the Supplementary Material we present regression tables showing that these results are valid for a twenty year period (1985-2005), two ten-year periods or four five-year periods, and that it is robust to the inclusion of other control variables such as individual country dummies (to capture any time-invariant country characteristic) and outperforms other indicators used to measure the productive structure of a country such as the Hirschman-Herfindahl [12,13] index and entropy measures [14]. A graphical example of this relationship is presented in Figure 3 f, which compares the growth predicted from the linear regression described by equation (6) and that observed empirically for the 1985-2005 period and *N=18*.

Finally, we show that the evolution of $M_{cp}$ exhibits strong path dependence, meaning that we can anticipate some properties a country's future new exports based on its current productive structure. This observation is consistent with the existence of an unobservable capability space that evolves gradually, as the ability of a country to produce a new product is limited to combinations of the capabilities it initially possesses plus any new capabilities it will accumulate. Countries with many capabilities will be able to combine new capabilities with a wide set of existing capabilities, resulting in new products of higher complexity than those of countries with few capabilities, which will be limited by this fact.

We show this using data collected between 1992 and 2000[§] and consider as a country's new exports those items for which that country had an $RCA_{cp}<0.1$ in the year 1992 and an *RCA* larger or equal to 1 by the year 2000. Figure 4 shows that the level of diversification ($k_{c,0}$) of a country and the ubiquity of its exports ($k_{c,1}$), predicts the average ubiquity ($<k_{p,0}>$) of a country's new exports and the average level of diversification ($<k_{p,1}>$) of the countries that were hitherto exporting those products.

---
[§] We choose 1992 as our starting point as the end of the Soviet Union and the unification of Germany introduce large discontinuities in the number and identity of countries.

This result is related to the idea that the productive structure of countries evolves by spreading to "nearby" products in The Product Space [15,16,17], which is a projection of the bipartite network studied here in which pairs of products are connected based on the probability that they are exported by the same countries. This last set of results suggests that the proximity between products in the The Product Space is related to the similarity of the requisite capabilities that go into a product, as countries tend to jump into products that require capabilities that are similar to those required by the products they already export.

## Discussion

Understanding the increasingly large gaps in income per capita across countries is one of the eternal puzzles of development economics. The intuition is that complexity is at the root of the explanation, as argued by both Adam Smith [1] and the recent endogenous growth theories [2,3], yet empirical research has not been able to progress along these dimensions because of the absence of adequate measures of complexity. Instead, empirical research has emphasized the accumulation of a few highly aggregated factors of production, such as physical and human capital or general institutional measures, such as rule of law, disregarding their specificity and complementarity. In this paper we have presented a technique that uses available economic data to develop measures of the complexity of products and of countries, and showed that (i) these measures capture information about the complexity of the set of capabilities available in a country, (ii) are strongly correlated with income per capita, (iii) are predictive of future growth, and (iv) are predictive of the complexity of a country's future exports, making a strong empirical case that the level of development is indeed associated to the complexity of a country's economy.

This paper has not emphasized the process through which countries accumulate capabilities, but has instead focused on their measurement and consequences. However, the results presented here suggest that changes in a country's productive structure can be understood as a combination of two processes (i) that by which countries find new products as yet unexplored combinations of capabilities they already have, and (ii) the process by which countries accumulate new capabilities and combine them with other previously available capabilities to develop yet more products.

A possible explanation for the connection between economic complexity and growth is that countries that are below the income expected from their capability endowment have yet to develop all the products that are feasible with their existing capabilities. We can expect such countries to be able to grow more quickly, relative to those countries that can only grow by accumulating new capabilities.

This perspective also suggests that the incentive to accumulate capabilities would depend, among other things, on the expected demand that new capabilities would face, and this would depend on how new capabilities can complement existing ones to create new products. This opens up an avenue for further research on the dynamics of product and capability accumulation.

Development economics has tended to disregard the search for detailed capabilities and their patterns of complementarity, hoping that aggregate measures of physical capital (e.g., measured in dollars) or human capital (e.g., measured in years of schooling) would provide enough guidance for policy. Our line of research would justify and provide guidance to development strategies that look to promote products (or capabilities) as a way to create incentives to accumulate capabilities (or develop new products) that could themselves encourage the further co-evolution of new products and capabilities, echoing ideas put forward by Albert Hirschman[18] more than 50 years ago, but adding the capacity to analyze them in practice.

### Acknowledgments:
We would like to thank M. Andrews, A.-L. Barabási, B. Klinger, M. Kremer, N. Nunn, L. Pritchett, R. Rigobon, D. Rodrik, M. Yildirim, R. Zeckhauser, participants at the Center for International Development's (CID) Seminar on Economic Policy and the Harvard Kennedy School Faculty Seminar, members of the CCNR at Northeastern University and the Ratatouille Seminar Series. We acknowledge support from the Growth Lab and the Empowerment Lab at CID.

**FIGURE CAPTIONS**

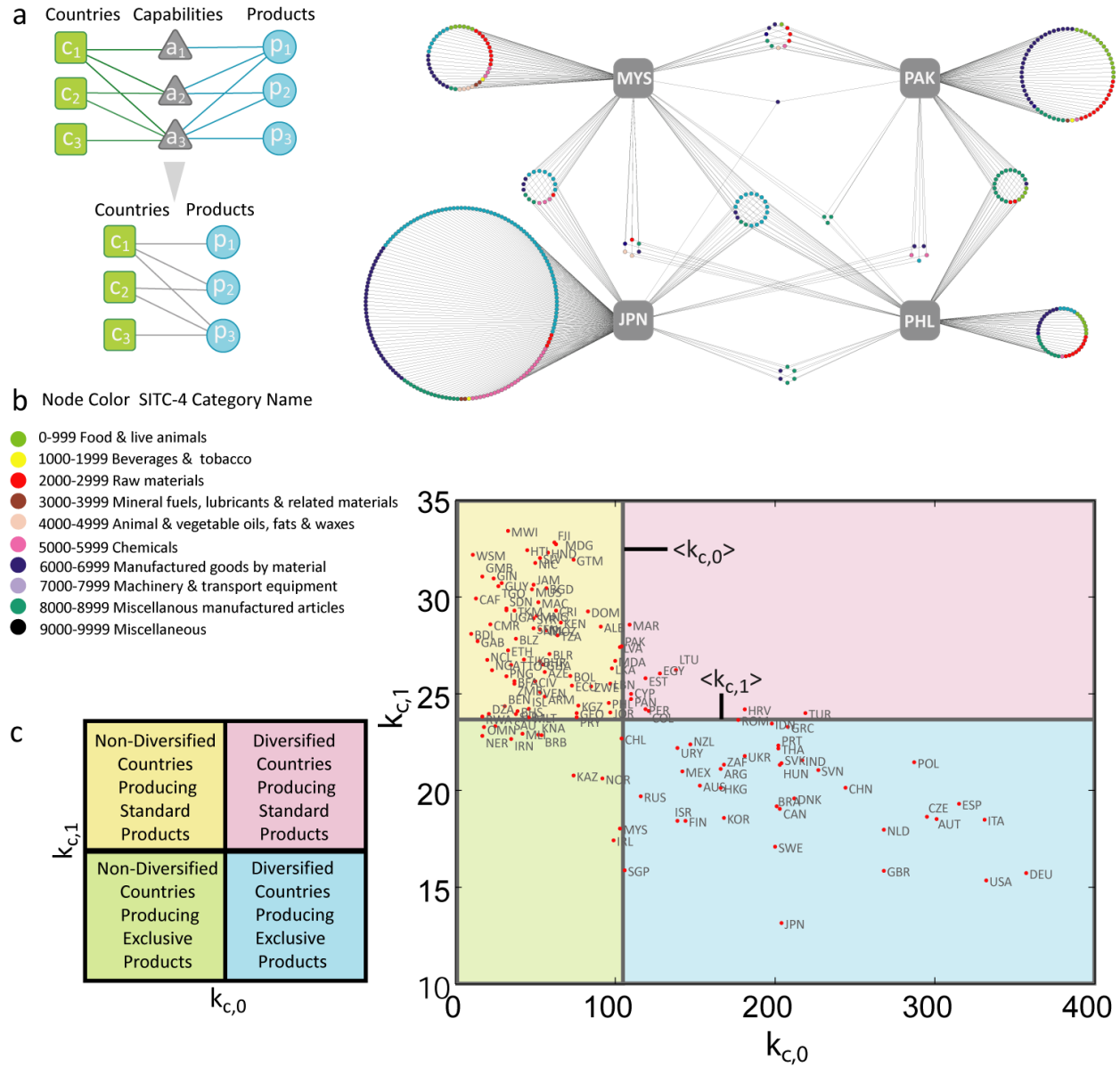

**Fig 1** Quantifying countries' economic complexity. **a,** A country will be able to produce a product if it has all the available capabilities, hence the bipartite network connecting countries to products is a result of the tripartite network connecting countries to their available capabilities and products to the capabilities they require. **b**, Network visualization of a subset of $M_{cp}$ in which we show Malaysia (MYS), Pakistan (PAK), Philippines (PHL), Japan (JPN) and all the products exported by them in the year 2000 (colored circles), illustrating how countries and products are connected in $M_{cp}$. **c.** $k_{c,0}$-$k_{c,1}$ diagram divided into four quadrants defined by the empirically observed averages $<k_{c,0}>$ and $<k_{c,1}>$

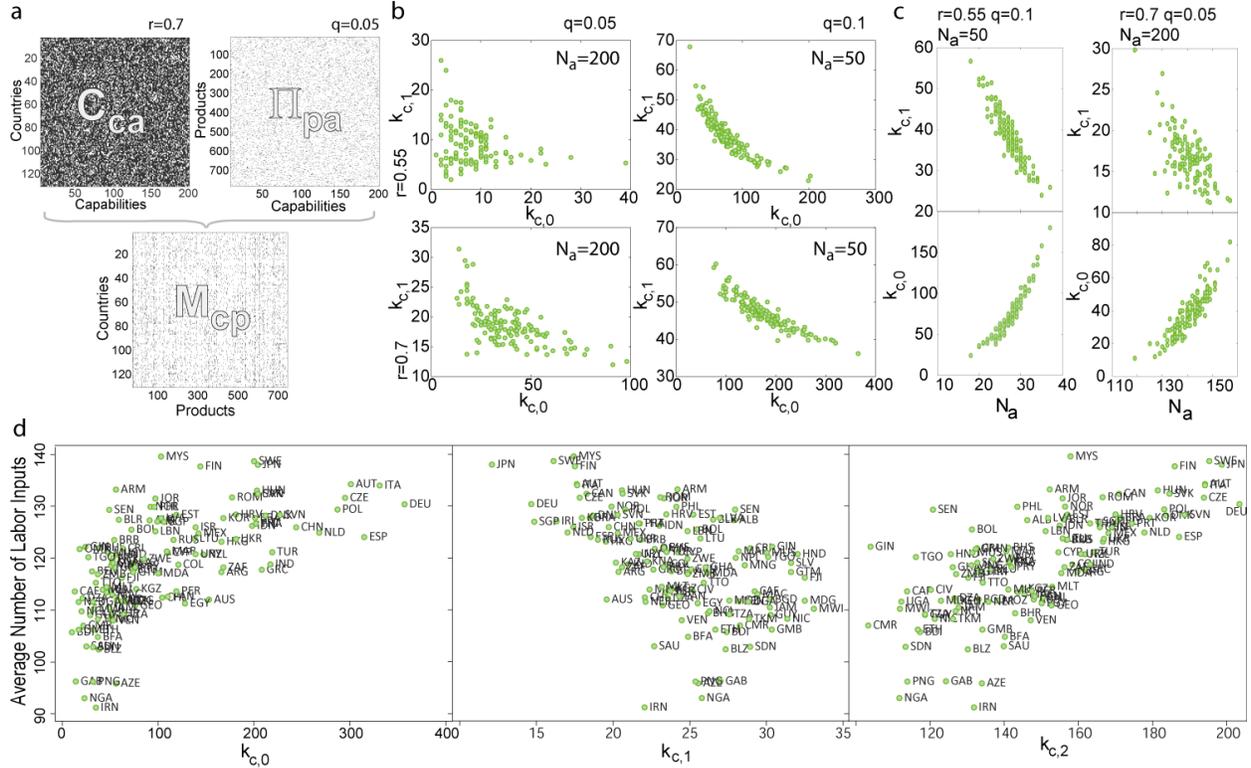

**Fig 2** Capabilities and bipartite network structure. **a,** We model the structure of $M_{cp}$ by taking two random matrices representing the availability of capabilities in a country and the requirement of capabilities by products and consider that countries are able to produce products if they have all the required capabilities. **b,** The $k_{c,0}$-$k_{c,1}$ diagrams that emerge from four implementations of the model described in **a**. **c,** $k_{c,0}$ and $k_{c,1}$ as a function of the number of capabilities ($N_c$) available in countries for two implementations of the model. **d,** Average number of labor inputs required by products produced in a country as a function of our first three components of $\vec{k}_c$.

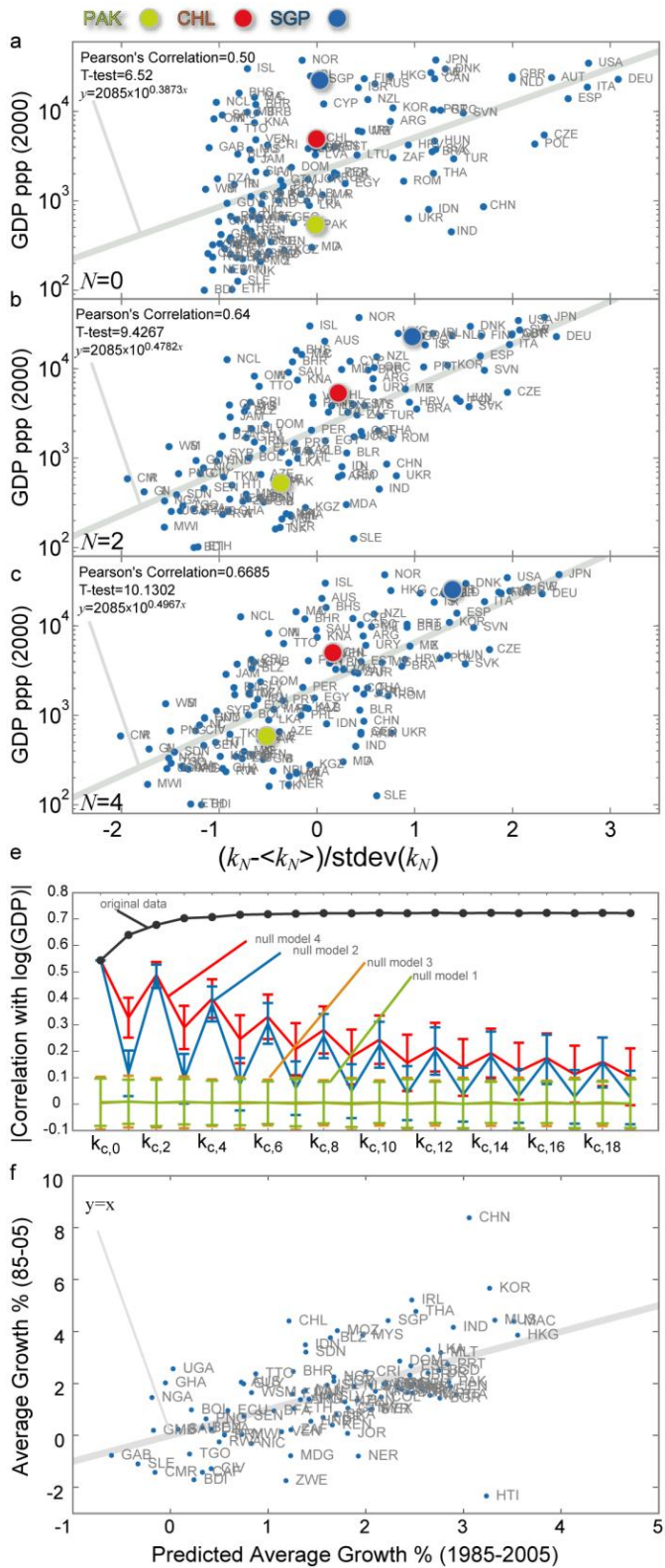
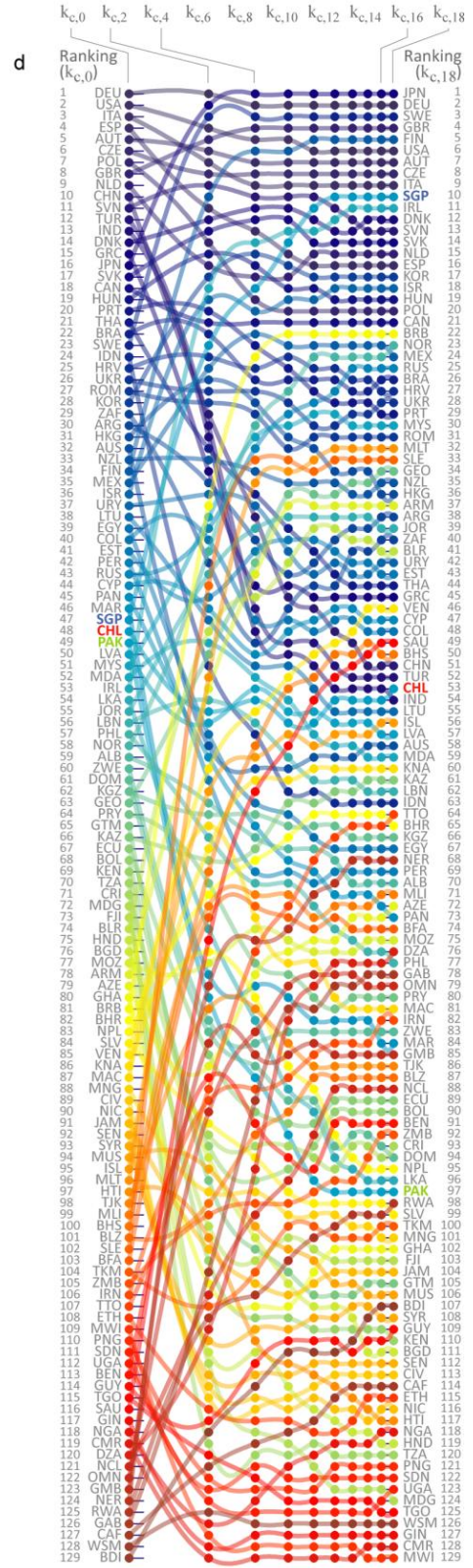

**Fig 3** Bipartite network structure and income (all GDPs have been adjusted by Purchasing Power Parity PPP. All figures were constructed with data from the year 2000). **a-c,** GDP as a function of our first three measures of diversification ($k_{c,0}, k_{c,2}, k_{c,4}$), normalized by subtracting their respective means ($\langle k_{c,N} \rangle$) and dividing them by their standard deviations (stdev($k_{c,N}$)). **a,** $k_{c,0}$, **b,** $k_{c,2}$. **c,** $k_{c,4}$. Comparison between the ranking of countries based on successive measures of diversification ($k_{c,2N}$) **e,** Absolute value of the Pearson correlation between the log(GDP) of countries and its local network structure characterized by $k_{c,N}$. **f,** Growth in GDP observed between 1985-2005 as a function of growth predicted from $k_{c,18}$ and $k_{c,19}$ measured in 1985 and controlling for GDP in 1985.

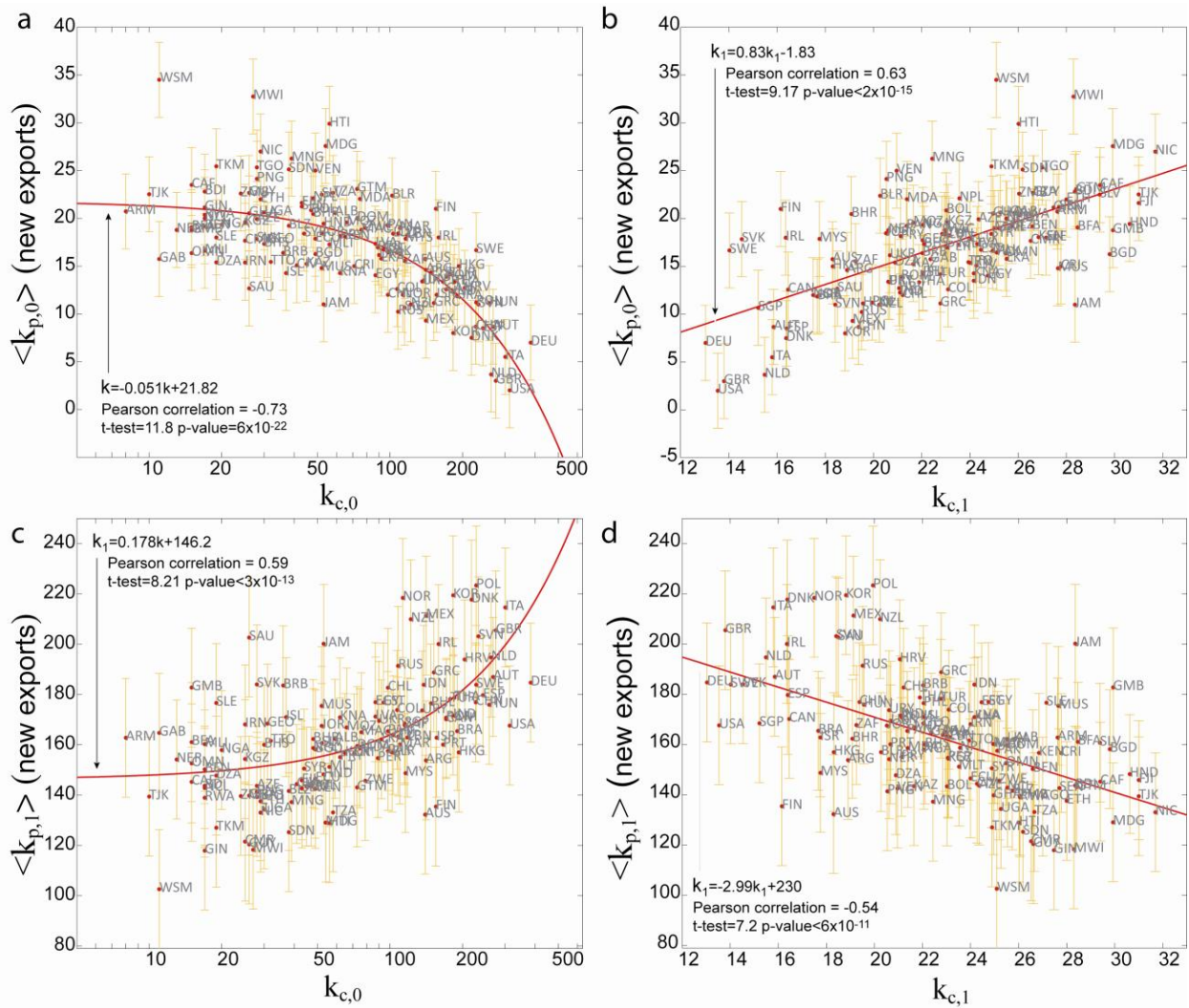

**Fig 4** Path dependent development. Average network properties ($\langle k_{p,0} \rangle$, $\langle k_{p,1} \rangle$) of the new exports developed by a country between 1992 and 2000 as a function of the diversification of a country $k_{c,0}$ and the average ubiquity of its products $k_{c,1}$ measured in 1992. **a,** $k_{c,0}$ v/s $\langle k_{p,0} \rangle$, **b,** $k_{c,1}$ v/s $\langle k_{p,0} \rangle$ **c** $k_{c,0}$ v/s $\langle k_{p,1} \rangle$ **d** $k_{c,1}$ v/s $\langle k_{p,1} \rangle$